\documentclass[10pt,conference]{IEEEtran}

\usepackage{hyperref}      
\usepackage{breakurl}      
\usepackage{color}         
\usepackage{graphicx}      
\usepackage{amsmath}       
\usepackage{amssymb}       
\usepackage{bytefield}     
\usepackage{algorithm}     
\usepackage{algpseudocode} 
\usepackage[table]{xcolor} 

\hypersetup{pdfauthor={{Louis-Emile Ploix,             
            Alasdair Armstrong, Tom Melham,
            Ray Lin, Haolong Wang, Anastasia Courtney}},
            pdftitle={Comprehensive Formal Verification
                      of Observational Correctness
                      for the CHERIoT-Ibex Processor},
            colorlinks=true,                          
            linkcolor=blue,                           
            citecolor=blue,                           
            urlcolor=blue                             
}

\title{
  Comprehensive Formal Verification\\
      of Observational Correctness\\
      for the CHERIoT-Ibex Processor}

\date{\tfm{Insert hard-coded date here}}

\author{\IEEEauthorblockN{Louis-Emile Ploix\IEEEauthorrefmark{1}, Alasdair
        Armstrong\IEEEauthorrefmark{2}, Tom
        Melham\IEEEauthorrefmark{1}, Ray Lin\IEEEauthorrefmark{1}, Haolong Wang\IEEEauthorrefmark{1}, and Anastasia Courtney\IEEEauthorrefmark{1}}\\
  \IEEEauthorblockA{\IEEEauthorrefmark{1}Department of Computer Science,
  University of Oxford} \IEEEauthorblockA{\IEEEauthorrefmark{2}Department of
    Computer Science and Technology, University of Cambridge}}

\begin{document}

\maketitle

\begin{abstract}
  The CHERI architecture equips conventional RISC ISAs with significant
  architectural extensions that provide a hardware-enforced mechanism for
  memory protection and software compartmentalisation. Architectural
  \textit{capabilities} replace conventional integer pointers with memory
  addresses bound to permissions constraining their use.  We present the
  first comprehensive formal verification of a capability extended \mbox{RISC-V} processor
  with internally `compressed' capabilities
  --- a concise encoding of capabilities with
  some resemblance to floating point number
  representations.

  The reference model for RTL correctness is a minor variant of the full
  and definitive ISA description written in the Sail ISA specification
  language. This is made accessible to formal verification tools by a
  prototype flow for translation of Sail into SystemVerilog.
  Our verification demonstrates a methodology for establishing that the
  processor always produces a stream of interactions with memory that is
  identical to that specified in Sail, when started in the same initial
  state.  We additionally establish liveness. This abstract,
  microarchitecture-independent \textit{observational correctness} property
  provides a comprehensive and clear assurance of functional correctness
  for the CHERIoT-Ibex processor's observable interactions with memory.
\end{abstract}

\section{Introduction}

Despite being known, researched, and partially addressed by various
technical measures for almost half a century~\cite{Memhist}, memory error
exploitations remain widespread and highly detrimental to computer
security. In 2019, Microsoft reported that 70\% of all security
vulnerabilities were caused by memory safety issues~\cite{MS2019}, and
out-of-bounds write is ranked number one in the SANS CWE top 25 most dangerous
software errors~\cite{CweSans}.  The issue is particularly acute for IoT
applications; like many embedded systems, these are often developed
in memory unsafe languages, such as C or C++.

CHERI~\cite{cherimips, introduction-to-cheri} is an approach to enforcing
memory safety that extends conventional ISAs with hardware-supported
\textit{architectural capabilities} for fine-grained memory protection and
scalable software compartmentalisation. A capability can be seen as a
memory address---the hardware representation of a software pointer---that
carries additional meta-data, notably the region of memory its address may
access and certain permissions constraining the capability's use. A correct
hardware implementation of an ISA with CHERI extensions ensures that no
memory accesses occur outside the permitted regions and that the
permissions are always respected.

The CHERI architecture further guarantees that capabilities cannot be
created arbitrarily, but only derived from other capabilities. The system
starts with certain \textit{root} capabilities from which all other
capabilities are derived---whether by a trusted loader, the OS, a
capability-aware compiler, or application code. Moreover, a new capability
can be derived from an existing one only by narrowing the region of memory
it can access or removing permissions. This crucial \textit{non-increasing
  monotonicity} property is enforced by the hardware and is what lays the
solid foundation for strong memory protection and software
compartmentalisation.

Early designs for CHERI processors had high memory overhead and memory
bandwidth consumption because the upper and lower bounds on the accessible
memory region were each represented with the same number of bits as the
address~\cite{cherimips}.  This has been replaced by a sophisticated scheme
of \textit{compressed} capabilities, greatly improving the practicality of
the approach~\cite{CHERIConcentrate}. This optimisation comes at the cost
of making certain address and bounds combinations unrepresentable, and a
requirement for the microarchitecture to implement efficient handling of
compressed capabilities and representability checks.

Processors that implement CHERI extensions to RISC-V are an especially
attractive target for hardware formal verification research. The
correctness of the RTL for a CHERI processor is, of course, the
indispensable foundation for the security guarantees of the CHERI
approach. Full and authoritative ISA specifications written in
Sail~\cite{Sail} are already available ~\cite{cheri-risc-v,cheriot-sail}.
In some cases, Sail was used for the ISA design. The
availability of a full, well exercised, and authenticated ISA specification
eliminates a steep practical barrier to applying formal verification.

Finally, the significant novelty of a CHERI ISA extension and the
microarchitectural optimisation of its implementation, including handling
of compressed capabilities, presents opportunities both to uncover bugs and
drive developments in methodology and tooling. The RISC-V base architecture
of a CHERI RISC-V processor also presents a relatively tractable---though
by no means trivial---formal verification objective, where one might
reasonably aim for more comprehensive results than only bounded properties.

\subsection{Contributions}

In this work, we demonstrate a practical methodology, with associated
proof-engineering and tooling, that achieves a comprehensive formal
verification of a RISC-V processor with CHERI extensions and
microarchitectural optimisations for internal compressed capabilities. Our
top-level correctness criterion---which we provisionally call
\textit{observational correctness}---is that the stream of memory
interactions engaged in by the processor will always be identical to that
stipulated by the authoritative ISA specification in Sail, when started in
the same initial state. We also verify liveness for these interactions
and---separately---correctness of the processor's instruction fetch logic.

No formal verification exercise should claim to establish deployed
processor `correctness' in an absolute sense. Our formal verification also
has some limits, and we delineate its technical coverage in
Sec.~\ref{sec:coverage}. We do however believe that the methodology demonstrated here
is both practical and provides significant assurance that a large class of
functional errors are absent from the design, relative to its
specification. Our correctness criterion is also clear and easy to
communicate at the abstract level of the ISA.

We demonstrate our methodology on Microsoft's open-source CHERIoT-Ibex
processor~\cite{amar2023cheriot} using the Jasper formal verification
environment from Cadence~\cite{jasper}. To the best of our knowledge, this
is the first formal verification effort with substantial coverage of a CHERI
processor that implements compressed
capabilities within the core. Our work revealed around 30 bugs in the
design, most to do with capability extensions. These are, of course, the
novel part of the architecture.  Four of these break the critical
monotonicity property of CHERI, the foundation for its utility in
preventing memory error exploitation.

In addition to demonstrating our methodology, our verification has
practical value in itself. The CHERIoT-Ibex RTL we have verified is, for
example, being used in the Sonata board for embedded systems being designed
by lowRISC~\cite{Sonata}.

Separately, we have also verified vanilla Ibex~\cite{ibex-verification} in the same
way and in the same configuration, with the only differences being the embedded
(RV32E) parameter is unset and physical memory protection (PMP~\cite{pmp})
is present---including PMP Enhancements for Memory Access and Execution
Prevention on Machine Mode (Smepmp/ePMP~\cite{smepmp}).
One bug in PMP checking for instruction fetch was found in vanilla
Ibex. This report, however, focusses largely on the CHERIoT-Ibex
verification.

This applied formal verification research has also driven the development
of a prototype compiler from Sail to a SystemVerilog reference model. This
has allowed us to use leading-edge commercial formal verification tools to
verify CHERIoT-Ibex against Microsoft's comprehensive ISA
specification~\cite{cheriot-sail} expressed in Sail. This novel flow
decisively overcomes the severe barrier to productivity reported in
previous CHERI verification efforts that relied on manual translation from
Sail to SystemVerilog~\cite{Gao:2021:EFV}.  More generally, this prototype
compiler makes the comprehensive, formal, and published ISA specifications
that exist in Sail directly available for industrial RISC-V processor
verification.

Our hope is that the methodology demonstrated here can be adapted for other
CHERI cores, and perhaps also for \mbox{RISC-V} processors in general. We
therefore make this preliminary report on our research available as a
resource for similar verification exercises. The open-source formal
verification code for the CHERIoT-Ibex processor~\cite{cheriot-ibex} is available
at~\cite{cheriot-ibex-verification}.  The open-source formal verification code for
the vanilla Ibex processor~\cite{ibex} is available at~\cite{ibex-verification}.

\section{Background on CHERI and CHERIoT-Ibex}\label{sec:background}

CHERI is a comprehensive systems architecture and security initiative, and
its research and development covers entire system stacks---from ISA design
up to operating systems and through to applications. The introductory
report~\cite{introduction-to-cheri} and CHERI architecture
report~\cite{cheriisav9} provide an overview and detailed specification of
CHERI respectively. Development practice in the CHERI project is to
concurrently co-design architecture extensions and their formal
semantics, their hardware implementations, and the software changes to
exploit them.

There have been several FPGA prototype ISA implementations~\cite{Picollo,
  Flute, Toooba}, as well as Morello~\cite{ArmMorello, Morello}, a
CHERI-extended A-class processor prototype developed by Arm and released in
January 2022.  The research reported in this paper addresses formal
verification of the RTL for CHERI processor designs. Our aim is to
establish confidence in the hardware foundation of CHERI security by
assuring the correctness of the lowest level implementation of the ISA,
including its microarchitectural optimisations.

Architectural \textit{capabilities} are the core concept of CHERI.  These
replace conventional integer pointers with memory addresses that are
integrally bound to meta-data that constrain their use~\cite{cheriot-sail, cheri-risc-v}:

\smallskip

\begin{itemize}
	\item \textbf{Validity tag.} A 1-bit, non-forgeable external tag
    bound to the capability. The tag indicates the capability’s
    approval for use. If a capability is invalid, it becomes simply a
    piece of data and cannot be used to access memory, serve as the
    target of a jump instruction, or produce any other operative
    effect.
  \item \textbf{Bounds.} The upper (`top') and lower (`base') bounds
    of a range in the memory space within which the capability is
    authorised for memory access operations, such as load, store, and
    instruction fetch. To save space, these fields are each shorter than a full integer
    memory address using an encoding scheme for bounds `compression'~\cite{CHERIConcentrate}
    that has some resemblance to floating point number
    representations.
  \item \textbf{Permissions.}  These define which types of operations
    involving the capability are allowed to run on the processor---including data
    memory read or write, instruction fetch (execute), and others.
  \item \textbf{Object type.}  Capabilities can carry an `object type', and can be `sealed' with that
  type. Sealed capabilities cannot be modified or dereferenced, but unlike
  invalid capabilities, they can be unsealed by executing other instructions
  that have authorisation to do this.
\end{itemize}

\smallskip

The capability-aware instructions in a CHERI ISA extension perform a range
of operations involving capabilities, including memory read/write,
capability field retrieval and manipulation, flow control, and so on. All
of these operations are constrained by the security principles of
CHERI---with violations raising exceptions (traps) or resulting in a
capability that has its validity tag cleared. It is important to note that
the address of a capability may stray outside its bounds. The bounds check
occurs only when the capability is used to access memory.

A valid capability can be created only from an existing valid capability,
and the principle of monotonicity prescribes that the privilege of a
derived capability cannot grow. Thus, bounds and permissions can only
shrink (or remain the same) throughout program execution, preventing
privilege escalation~\cite{cheriisav9}.

In a RISC-V processor with CHERI extensions, certain architectural
modifications are made to support CHERI's features: the general-purpose and CSR
register files are extended to contain full capabilities and memory is tagged
with one bit for every unit of memory that can contain a capability.
Some new CSRs are also added. The PCC, for instance, is
introduced as a special capability to hold the bounds of the currently
executing code.  Extending RISC-V for CHERI generally minimises the
influence of these modifications on the processor microarchitecture,
keeping the overall design (e.g. pipeline structure and memory hierarchy) largely
intact~\cite{introduction-to-cheri}.

\subsection{CHERIoT and CHERIoT-Ibex}\label{sec:corr}

CHERIoT is a reduced version of CHERI architected specifically for 32-bit embedded
devices, designed by Microsoft~\cite{amar2023cheriot}.  Although the
fundamental ideas remain the same, the CHERIoT ISA differs slightly from the
original CHERI ISA to better fit its embedded design purpose.

CHERIoT-Ibex~\cite{cheriot-ibex} is an open source fork of the lowRISC Ibex~\cite{ibex}
embedded processor, written in SystemVerilog and designed and maintained by
Microsoft research. It extends Ibex with the CHERIoT variant of CHERI, as
specified in Sail by the team at Microsoft. Like Ibex, its default
configuration has a three stage pipeline, no branch predictor, no virtual
memory and no (provided) cache. It allows misaligned memory accesses and
supports compressed RISC-V instructions. 

CHERIoT adopts a capability encoding and compression scheme similar to
64-bit CHERI Concentrate~\cite{CHERIConcentrate}, which keeps both the
permissions and the bounds of each capability in compressed
form. Permissions are compressed so that only useful combinations are
represented. For example, no capability will ever need to both write and
execute. This means that three \textit{root capabilities} are required in
CHERIoT: one for execution, one for memory read/write, and one for
sealing/unsealing.

Capability Bounds Compression exploits address redundancy to form a
representation for bounds similar to floating point. A capability has four
fields that are used to obtain the bounds of the capability. The
\textit{address} is the usual 32-bit field and can store any data. Two
9-bit fields $B$ and $T$ are used to derive the middle bits
of the \textit{base} and \textit{top} bounds, respectively, and an
\textit{exponent} field $E$ is used as a shift. It can take values 0--14 or 24,
with the latter chosen in order to represent the entire address space. It is itself
stored in a compressed form.

The following pseudocode, adapted from the Sail specification~\cite{cheriot-sail},
demonstrates the decompression algorithm:

\begin{algorithm}
  \caption{Capability Bounds Decompression Algorithm}\label{alg:cap}
  \begin{algorithmic}
    \Require $a, B, T, E$
    \Ensure $b, t$ are the 32 and 33 bit base and top bounds
    \State $a_\text{mid} \gets (a \gg E) ~ \& ~ 0\text{x}1FF$
    \State $a_\text{hi} \gets 1 ~ \textbf{If} ~ a_\text{mid} <_\text{unsigned} B ~ \textbf{Else} ~ 0$
    \State $t_\text{hi} \gets 1 ~ \textbf{If} ~ T <_\text{unsigned} B ~ \textbf{Else} ~ 0$
    
    \State $c_b \gets -a_\text{hi}$ 
    \State $c_t \gets t_\text{hi} - a_\text{hi}$ 
    
    \State $a_\text{top} \gets a \gg (E + 9)$
    \State $b \gets (((a_\text{top} + c_b) \ll 9) ~ | ~ B) \ll E$
    \State $t \gets (((a_\text{top} + c_t) \ll 9) ~ | ~ T) \ll E$
  \end{algorithmic}
\end{algorithm}

\noindent In the RTL for CHERIoT-Ibex, the correction terms named $c_b$ and
$c_t$ in this code are called \verb|base_cor| and \verb|top_cor|
respectively. As will be discussed in Sec.~\ref{dti:sec}, they are cached
in the microarchitecture for CHERIoT-Ibex, requiring some work to establish
a data-type invariant.

A fundamental consequence of bounds compression is that not all
combinations of addresses and bounds are \textit{representable} in a
capability. There are consequently numerous checks for representability in
the Sail specification of the ISA. Realising these semantics in the RTL
implementation can't be done by simply copying the Sail code, but requires careful design
of optimised microarchitecture for performance.

\section{Methodology Overview}\label{sec:methodology}

In this section, we give an overview of our approach and methodology, with an
outline of the structure of our verification.  We begin with some
principles.

First, our work on CHERIoT-Ibex aims for verification coverage that is as
comprehensive as possible. All our SVA properties have therefore been
proved to hold to an unbounded depth, as only incomplete coverage is
given by bounded proofs. This is a challenging but reasonable
ambition because of the small size and simplicity of the RISC-V core.
Most of our unbounded proofs have been obtained by $k$-induction, and
achieving this aim has necessitated the development of numerous invariants
at important points within the microarchitecture.

We have also put in significant effort to make regression fast. This, of
course, aids productivity---for example when bug-fixes are checked or more
basic changes are made, which is important for an open-source core. It also
gives some hope that the verification might be re-run in non-commercial
verification tools. Internal invariants are again the key here.

Our invariants are likely to be too microarchitecture-specific to be
directly usable in another verification project. They may, however, be
helpful examples of the sort of properties that may be required in future
efforts on more challenging CHERI or RISC-V ISA implementations.

More informative and potentially adaptable is the overall structure of our
verification. Presented top-down, a sketch of this is as follows:

\begin{enumerate}
  \item At the top level, perform a structural decomposition into
    instruction fetch and the execution
    pipeline~(\ref{trace-equivalence}). The Sail ISA specification does
    not model (nor aim to model) an instruction cache, so these two components
    are verified independently.
  \item \label{trace-equivalence} For the execution pipeline, prove that
    the infinite stream of interactions with memory produced by the
    execution pipeline is identical to that specified in Sail, when both
    are started in the same initial state. This top-level property
    expresses observational correctness.  Proving it requires a certain
    state-matching property (\ref{state-matching}), some top-level memory
    properties~(\ref{memory}), and liveness (\ref{liveness}).
  \item \label{state-matching} State-matching follows from a collection of
    what we call `continuity' properties (\ref{continuity}). It establishes
    that the processor architectural state is equivalent to that of the
    iterated specification at every point in a certain infinite series of
    points in time.
  \item \label{continuity} Continuity properties establish that the processor
    state does not change between instruction executions. They follow from
    a collection of end-to-end functional correctness properties for the
    instructions of the ISA~(\ref{end-to-end}), together with a number of
    internal invariants~(\ref{invariant}). Specifically, they verify that each time the
    specification is queried, the inputs to the specification match the
    old outputs from the specification.
  \item \label{memory} The top-level memory properties are an
    extension of the end-to-end instruction correctness
    properties~(\ref{end-to-end}), again supported by certain
    invariants~(\ref{invariant}). They establish that between matched pairs of states
    the processor and specification produce the same outputs to memory.
  \item \label{liveness} Liveness in our verification is under the
    assumption of bounded memory response times and that waited-for-interrupts
    arrive within bounded numbers of clock cycles. The bounds
    are low enough for the proof to be tractable, but high enough to
    encompass any period of execution that could lead up to a processor
    stall following a memory request or a wait-for-interrupt instruction.
  \item \label{end-to-end} The end-to-end instruction correctness
    properties methodologically constitute a case split over instruction
    types (where illegal instructions are considered one such type)
    and establish that each instruction entering the
    pipeline mutates the ISA state exactly as specified in Sail. This is
    the heart of the verification and also where most of the bugs were
    exposed.  Verifying end-to-end instruction correctness requires a
    number of helper invariants, including the data-type invariant about
    compressed capabilities alluded to in Sec.~\ref{sec:background}.
  \item \label{invariant} The invariants are almost entirely discovered iteratively and
    manually by observing counterexample traces of failed $k$-induction
    proofs. In Cadence terminology, this is called `State Space Tunnelling'.
\end{enumerate}

Many of the details and components of this methodology are
well-known, and we make no claims to novelty for them. End-to-end
instruction verification, for example, has been compellingly demonstrated
by Reid in~\cite{Reid-2016-EEV}, though our work differs in a number of
significant aspects. Case splitting over instructions is
standard. Our workhorse model-checking algorithm, $k$-induction, is widely
used---and advanced industrial verification engineers will be all too
familiar with the invariant discovery process.

Our aim here, however, is to demonstrate the structure of a complete
methodology that achieves a novel and significant verification result---the
observational correctness, with respect to the authoritative Sail
specification, of an industrial (and open source) RISC-V processor
implementation for the innovative CHERI instruction set architecture.

\section{Integrating the Sail Specification}

Sail~\cite{Sail} is a first-order imperative language for describing instruction set architecture
semantics that has been used to specify various architectures, including
RISC-V~\cite{sail-risc-v}, CHERI RISC-V~\cite{cheri-risc-v}, and the CHERIoT
extension to Ibex~\cite{cheriot-sail}. One of the goals of Sail is to provide a
language that can be transformed into other representations to support
various use cases. For example, Sail can generate formal definitions in
deductive theorem provers (Isabelle, HOL4, or Coq) or create a sequential
emulator for testing.

These established Sail tools are targeted at reasoning about formal properties
of a Sail specification, at concurrency modelling, or at emulation,
validation, and testing of machine code. To date, there has not been a flow
in the standard Sail environment that integrates Sail as a specification
language for formal hardware verification methodologies with state of the
art model checking tools, which work with more conventional languages such
as SystemVerilog and VHDL.

\subsection{Compilation of Sail Specifications to SystemVerilog}

Without an existing SystemVerilog specification for CHERIoT-Ibex,
our options for this research were to write a new
specification by hand, in which we translate the Sail into SystemVerilog
manually, or to automatically convert an existing specification into
SystemVerilog. Previous experience has shown that writing a new
specification can be a bottle-neck for verification
efforts~\cite{Gao:2021:EFV}, especially if one wants to cover all the
existing RISC-V instructions in addition to the CHERI features.  Writing
any large specification from scratch would also invariably entail a
significant validation effort to avoid or fix specification bugs. For these
reasons, we chose to develop a prototype conversion tool from the existing
Sail CHERIoT-Ibex specification to SystemVerilog.

In 2020 the Sail RISC-V model was selected as an official executable
golden model by RISC-V international, with a working group to oversee
development of the model. This means that currently, task groups
producing new RISC-V extensions are required to extend the Sail model
with their extension for ratification (although in practice, waivers
were given in the past for certain extensions, such as hypervisor).

The primary challenge in this work is that Sail supports specifications
that are parametric over bitvector widths in ways that are not easily
representable in SystemVerilog. Specifically, bitvector widths can depend
on runtime values and are statically checked using refinement
types~\cite{Vazou2013,Rondon2008}. But, for our prototype
implementation, there already existed both various monomorphisation
passes~\cite{Sail} and a Sail-to-SMT conversion pipeline that maps Sail
expressions into the fixed-width bitvector fragment of
SMT~\cite{sail-ifetch}. We therefore used this SMT fragment as the
intermediate representation from which we generated SystemVerilog
expressions, with some small extensions whenever we needed to generate
a specific SystemVerilog construct without an SMT equivalent.

For the overall imperative structure of the code, we chose to generate
SystemVerilog functions, because the semantics closely matches the
imperative semantics of Sail.  But this is not ideal for typical formal
verification tools, and in future versions we intend to generate more
idiomatic SystemVerilog that makes extensive use of modules.

To facilitate smooth integration into the formal verification, we
encapsulated the generated SystemVerilog specification with a small API
layer on top. Since the specification is machine-compiled, it can make
reference to unwieldy names and might contain unused registers for
extensions we do not support, such as floating point or vector
registers. These registers can be omitted from the API and are internally
left as `free' undriven signals.

The generated SystemVerilog is purely combinational and monolithic. It maps
a given processor architectural state $S$ to the state that results from
running a given instruction on state $S$. We tested the resulting
SystemVerilog reference model by re-using the existing Sail test
suite. Overall, we did not encounter many bugs in the translation, but
further validation could be done to gain additional confidence. Of course
the top-level observational correctness result obtained in our work also
contributes to our confidence in the translation.

\subsection{Sail Specification Modifications}

Unfortunately, the CHERIoT-Ibex Sail specification as it stands is not
fully suitable for hardware formal verification. In just a few cases the
Sail specification requires behaviour that the English language RISC-V
specification allows but does not expect. See the EBreak instruction for
example~\cite{ebreak-english,ebreak-sail}, where the value stored in
the MTVAL CSR can be either 0 or the PC according to the RISC-V specification,
but the Sail would require it always be the PC.  The Sail specification also
presents some scalability issues for hardware formal verification purposes. This is all
generally a result of the fact that Sail has, up to this point, been used
almost exclusively in an `upward' fashion---for reasoning, emulation, or
testing on the foundation given by the ISA specification. For example,
security-related properties have been proved that follow from the
specification~\cite{cheriot-sail-properties} or the specification has been
used to generate tests~\cite{Sail}. To date, little work has gone into
proving that a hardware implementation conforms to that specification.

This all meant that the verification reported here is based on a modified
fork of that specification.  While we, the authors, are confident that our
specification is compliant with RISC-V, we must nonetheless acknowledge
that our statements are about conformance with our own specification, not
the original. In particular, the following key changes were made:

\begin{enumerate}
  \item Add support for Embedded (E) mode, which is used by CHERIoT-Ibex.
  \item Remove unused extensions (VEXT, for example).
  \item Make RVC instruction decompression a distinct step to execution. This is the only change
    made with the express purpose of making verification easier. This
    massively cuts elaboration times for the generated SystemVerilog model
    and the memory footprint of the specification.
  \item Relax some instruction decodings (i.e. successfully decode instruction bits which
    do not strictly match the pattern for any particular instruction). This is allowed by the RISC-V specification.
  \item Minor configurability improvements, such as choosing what MTVAL is set
    to on an EBREAK instruction, as mentioned above.
\end{enumerate}

Some additional changes to the Sail were needed for vanilla Ibex verification,
especially for PMP:

\begin{enumerate}
  \item Separate misaligned memory accesses into two 4 byte aligned regions
    for PMP checking. In Ibex, misaligned memory requests
    (i.e. those crossing a 4 byte boundary) are split into two separate requests,
    instead of one misaligned request. This distinction was important when verifying the PMP
    implementation of vanilla Ibex. The RISC-V specification does allow
    two aligned 4-byte accesses to be checked instead of the one.
    This difference is observable on region boundaries.
  \item Add Smepmp support using a specification
    written in an open pull request currently on the public sail-riscv
    repository~\cite{sail-risc-v-smepmp}.
    This is because upstream sail-riscv does not have support for Smepmp even though the specification was released as v1.0 in 2021~\cite{smepmp}.
\end{enumerate}

\section{Global Data-Type Invariant}\label{dti:sec}

Registers holding capabilities in CHERIoT-Ibex store them in compressed
form, but hardware operations on capabilities typically require them to be
decompressed. This is a relatively expensive computation, requiring
multiple comparisons, variable length shifts, and pointer width
additions. In particular, decompression requires the computation of the two
values \verb|top_cor| (two bits) and \verb|base_cor| (one bit), introduced above in
Sec.~\ref{sec:corr}. For performance, these are cached in CHERIoT-Ibex and
stored as additional fields with the compressed capability. Since these
fields are not cached in the Sail, proving that the instructions that
operate on them conform to the specification requires the cached
values---at the point of use---to be correct.

To ensure this, a processor-wide data-type invariant (DTI) was introduced
and proved. Later proofs of capability-manipulating instructions could then
rely on the DTI to constrain the state space. A virtually identical
data-type invariant was also proved in later work on CHERIoT-Ibex
verification with VeriCHERI~\cite{vericheri2024}. We discuss this
research in more detail in Sec.~\ref{sec:related}.

The data-type invariant is hard for typical model checking engines to
prove directly quickly. To address this, we developed a decomposition
methodology that allows the DTI to be established in around two minutes,
and which we subsequently also used in our verification of memory
properties and of PMP for vanilla Ibex.

Our methodology resembles what one would classically do in an interactive
theorem prover, but realised within a model-checking environment capable of
assume-guarantee reasoning. We prove each of the following, where steps
1--4 can be proved independently and step 5 is provable by $k$-induction
from 1--4.  Here `block' refers loosely to those same blocks, such as
EX or CSR, as seen in Fig.~\ref{follower.fig}. Some blocks are case split
for speed; in particular, the CSR block is split into individual blocks for
each CSR it contains.

\begin{enumerate}
  \item For each block with state, the capabilities stored at reset satisfy
    the DTI. 
  \item For each block with state, if each capability stored 
    satisfies the DTI and each input capability satisfies the DTI,
    then any output capability also satisfies the DTI.
  \item For each block with state, if each capability stored satisfies the
    DTI and the capabilities presented at the inputs satisfy the DTI, the
    capabilities stored within the state at the next clock cycle will
    satisfy the DTI.
  \item For each block without state,  if the capabilities being
    presented at the inputs satisfy the DTI, any output capabilities also
    satisfy the DTI.
  \item For every block with state, all of the capabilities stored always satisfy the
    DTI.
\end{enumerate}

\noindent Each of these properties prove almost instantly with
$k$-induction at a low value for $k$, 1--2, with the longest (the MTCC CSR case)
taking two minutes or so.

Establishing this data-type invariant was essential to the convergence of
many correctness properties. But it was also critical for the discovery of
bugs---as, generally speaking, any violation of the DTI was an indication
of incorrect handling of a capability somewhere. In several cases this
contributed towards a monotonicity-breaking bug that allowed for illegal
movement of the bounds of a capability.

As will be seen in Sec.~\ref{sec:memory}, the global data type invariant
was also extended to validate certain well-formedness assumptions about
capabilities coming from memory.

\section{End-to-End Instruction Correctness}\label{sec:end-to-end}

The core of our verification methodology is a case split over instruction
type, where each type is verified separately employing the so-called
end-to-end approach. In particular, we verify that the execution of each instruction
is correct with respect to the Sail specification, from the point immediately
after it has been fetched to when it has finished applying its effects to
architectural state. During this period, we observe the changes made to
architectural state and verify their correctness against
one iteration of the main loop of the Sail specification.

In CHERIoT-Ibex, a given instruction may mutate several different state
components as it moves through the pipeline, always to distinct blocks of
architectural state and often at different times.  For example, in
CHERIoT-Ibex an instruction updates the CSRs when it finishes the execute
stage, but it updates the register file when it finishes the writeback
stage. An instruction may also spend several clock cycles in the execute
stage of the pipeline---for example when the pipeline is stalled waiting for
a memory response for the instruction ahead of it.

Our verification set-up handles these timing behaviours as follows. In the
cycle that an instruction will continue from the execute stage to the
writeback stage we evaluate the Sail specification on the architectural state of
the implementation as it is at that point. When the clock flops,
we store these results in a \textit{pipeline follower}~\cite{Reid-2016-EEV}
for reference in subsequent stages of pipeline execution of that instruction.

\subsection{The Pipeline Follower}

\begin{figure*}[t]
  \begin{center}
    \includegraphics[width=140mm]{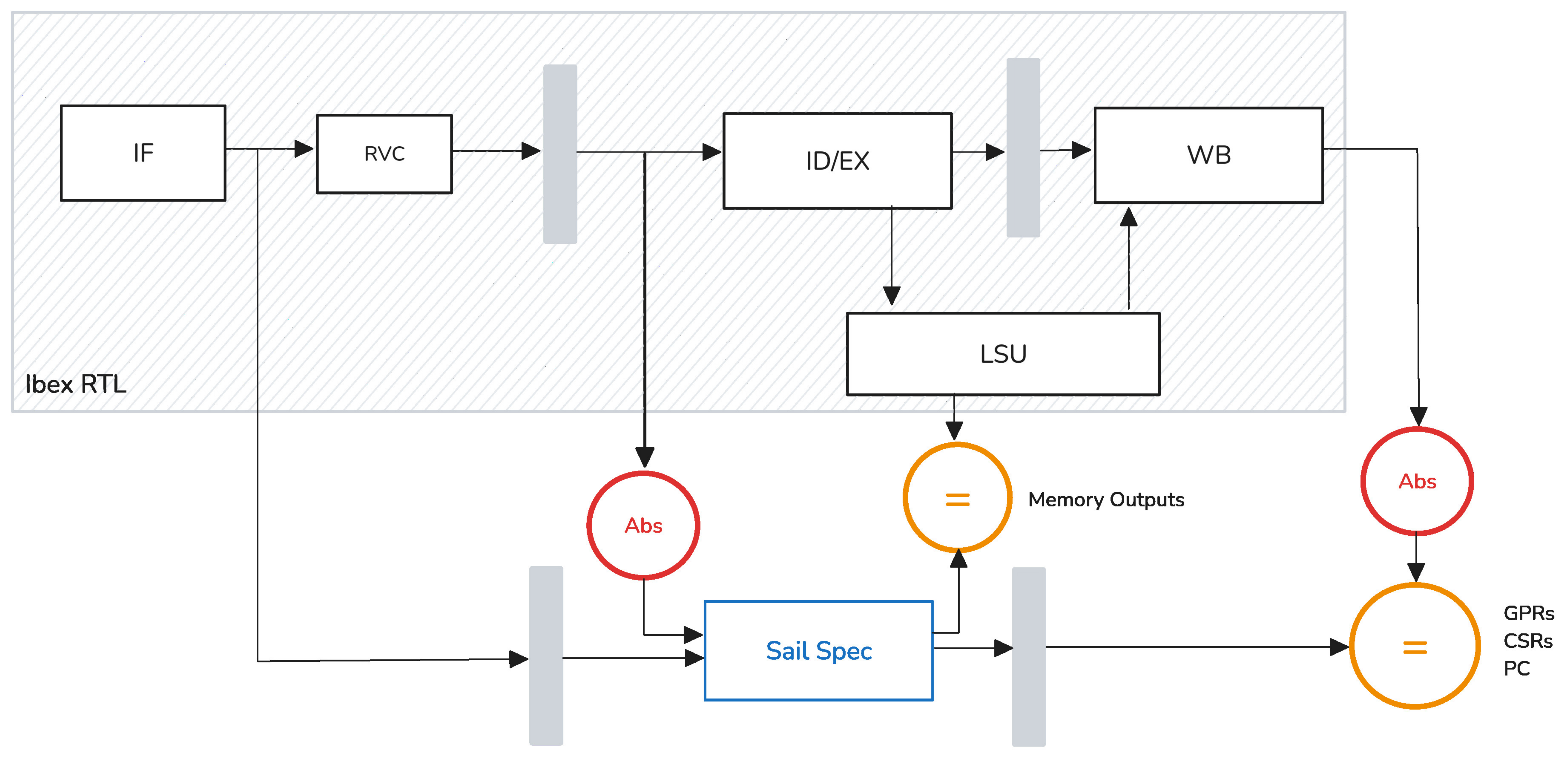}
  \end{center}
  \caption{CHERIoT-Ibex pipeline with specification installed using a pipeline follower.}\label{follower.fig}
\end{figure*}

The pipeline follower is a central component in our verification
infrastructure that steps the saved specification data forward alongside
the instruction, as it moves through the pipeline. When any update is made
to architectural state, it is checked against the stored specification
results immediately. Evaluating the specification at the boundary
between the execute stage and the writeback stage minimises the time
span that a property needs to consider to be proved. It also ensures
that all necessary information to run the instruction (e.g. load hazards)
is available, making the querying of the specification simpler and self
contained: i.e. this query of the specification does not rely on
the previous one.

In prior work (most notably ISA-Formal~\cite{Reid-2016-EEV}), the specification was queried
in the same cycle in which verification occurred (i.e. at the point of retirement),
with the post-state (loosely) abstracted from the results of the
stage preceding writeback and the pre-state (loosely) abstracted from the state presently held in writeback (e.g. the
current register file). While this is generally an oversimplification of what has been done by ARM,
this model does nonetheless not apply to CHERIoT-Ibex. CSRs, for instance, will be mutated in the
previous pipeline stage (execute) instead of after writeback, meaning by the time an instruction is retiring the
stored state in the processor is no longer the pre-state of that instruction.

For this reason we use an alternative strategy wherein we store, in the pipeline follower,
the result of evaluating the specification on the abstract pre-state of the processor
at the point the instruction begins executing.
Storing the entirety of the initial state in one cycle works because the CHERIoT-Ibex pipeline is short,
and does have all the necessary pre-state information available at this time
(namely at the end of the ID/EX stage). In practice, for longer pipelines sharing this quirk,
this may not be possible and pre-state would instead need to be
accumulated throughout each pipeline stage as it is first accessed. The specification could then be
queried either as soon as the pre-state is available (as we do), or at the point of retirement.
Since we evaluate the specification early we also get the benefit of being able to use its outputs for
case splitting and as helpers for defining invariants. This way we can also verify correctness of memory
requests as they happen, instead of storing the history of memory operations until retirement.

The pipeline follower is depicted in Fig.~\ref{follower.fig}.  Importantly,
the presence of the pipeline follower avoids assuming that only one instruction
goes through the pipeline at a time and check only that instruction, as with
Burch-Dill flushing~\cite{10.1007/3-540-58179-0_44}.  We instead allow the
implementation to run in any state that might exist during real world
execution.

In our approach, progress of the specification results through the pipeline
follower is controlled by the same stall/flush hardware logic---the same
control signals---as the implementation pipeline itself. This is an
important and general design decision in our methodology, which enables
tractable installation of the specification into a complex pipeline. It is
not necessary for the pipeline follower to implement `the same pipeline
stall/flush logic as the datapath'~\cite{Reid-2016-EEV}.

This approach greatly simplifies the construction of the pipeline follower
and installation of the specification. But, of course, using this
end-to-end approach on its own---without additional verification
properties---could miss pipeline control bugs. Similarly, bugs relating to
the forwarding logic, or changes made to state between two instructions,
may also be missed. In our methodology, therefore, we complement the
end-to-end instruction correctness proofs with what we call continuity
properties, which closes these gaps. We explain these in detail in
Sec.~\ref{sec:obs}.

\subsection{State Abstraction}

In order to meaningfully compare the results of the specification against
the implementation they need to be mapped into a common view of processor
`state'. In common with many other processor verification efforts, we
achieve this by implementing an abstract state API on top of the processor
RTL. This deals with three things:

\begin{enumerate}
  \item Determining the observed initial state of an instruction. This includes
  checking if the value of a register is forwarded or can be pulled straight from the register file.
  \item Determining the post-update state of each piece of architectural state
  at the time that piece of state is changing.
  \item In the case of a CHERI processor, mapping capabilities into a common type to that used in the
  specification. This is necessary due to the differences in their representation.
\end{enumerate}

RVFI~\cite{RVFI} provides an alternative means of getting trace information
from the processor.  We do make limited use of internal RVFI signals,
but we do not use the RVFI interface as a whole. RVFI does not fit our
end-to-end methodology, as it provides a limited view of the mutation
of architectural state, especially with respect to the CSRs.

\section{Memory}\label{sec:memory}

The top-level input ports on CHERIoT-Ibex that communicate data to and from
instruction and data memory must follow a well-defined protocol for how
quickly requests are allowed to be granted and how quickly responses may be
returned~\cite{ibex-mem-docs}.  In our verification, this protocol is
expressed as a straightforward collection of SystemVerilog
assumptions. These ensure that a response cannot come in the same cycle as
the associated grant, that grants only come when a request is being sent,
and that a response cannot come unless there is an outstanding request.

Subject to these constraints, the CHERIoT-Ibex memory protocol allows
requests and responses to be granted and received respectively after an
unbounded amount of time.  But proofs under unbounded liveness assumptions
are often computationally challenging. We address this in two ways. First,
we set a bound of 10 cycles for grants and responses to be returned.  This
makes a proof of processor liveness tractable, and it makes many small
helper properties prove within seconds by simply unrolling to that
bound. Second, we use proof strategies such as alternative LSUs; see
Sec.~\ref{src:memcorrectness} for details of this. Combined, these two
devices enable formal verification engines to obtain conclusive proofs of
the high level memory correctness properties we wish to have.

The other objective for the assumptions about memory in our verification
set-up is to ensure that both the specification and CHERIoT-Ibex receive
the same data when memory is read. This is the only way in which we
`interfere' with CHERIoT-Ibex, by choosing the values to provide on its top
level ports that receive data from memory.

The Sail specification is a purely combinational function that computes the
effect on architectural state of a complete instruction execution. For some
instructions, the resulting state updates will involve data that comes from
memory.  The specification therefore has input ports that represent any
data that will come from memory as a result of the current instruction
execution. These are primary inputs in the Verilog for the specification,
and so the data on these ports are encoded in the model checker as symbolic
values that we assume satisfy the constraints in
Sec.~\ref{src:memconstrain}.  Since the specification is evaluated in the
ID/EX pipeline stage, we store these symbolic values in the pipeline
follower so that in the future cycles in which CHERIoT-Ibex itself receives
data from memory, we can assert that the received data is the same as was
given to the specification.

\subsection{Memory Correctness Properties}\label{src:memcorrectness}

The memory properties in our verification are of two types: properties
about the memory requests the processor makes, and properties about the way
CHERIoT-Ibex uses the response. In the former case we are checking that
the correct addresses, \verb|wdata|, \verb|be|, and \verb|we|, are sent and
that the correct number of requests are made for the given instruction.
In the latter case we are checking that for load/store instructions the value
written to the register file (if any) is the correctly derived value from the response
from memory, and that no CSRs or the PC are broken along the way (essentially that
during the stall there are no architectural state changes to these registers).

Proving properties about request data is by no means trivial, but it
ultimately reduces to finding a system of invariants, one for each state of
the LSU (load-store unit), that are proved across each edge of the LSU
state graph in a similar sort of fashion to how the DTI was proved.  The
LSU is the Ibex RTL module that implements the state machine and logic for
handling memory operations. These properties are initially proved in the
case split, i.e. separately for loads and stores. They are later used as 
lemmas for the top-level memory properties, which prove the same properties
of any instruction, including non-memory instructions: for any
instruction the outputs to memory will match those of the specification.

Without further complexity management, the proofs about responses are
nonetheless slow to obtain. We solve this with a series of helper lemmas proved over
every cycle of the memory request. They are of the form `if the memory grant/response was
to arrive \textit{now}, the final result produced by the LSU to be sent to the WB stage
would be the correct one'.

Technically, this is done by introducing `alternative LSUs' into the RTL.
In our proof, we instantiate multiple times a modified version of the LSU:
one for before the first response from memory, one for after while we wait
for a second grant, and one for when we wait for a response in the WB stage.
In each instance, the alternative LSU's state is mostly just that
of the original LSU, but the final response is always assumed to be
arriving now. Since these alternative LSUs do not
contain any state of their own and do not drive any signals outside of
themselves, they provide a view, in any given clock cycle, of the
`alternative reality' in which a memory response was to arrive in that
cycle.  We can then assert that in this `alternative reality' the desired
property (that the value received from memory is processed correctly) is
true.

The model checker can chain these together and prove these properties about these additional LSUs
easily with $1$-induction, independent of the size of the memory response time bound,
and can leverage this into a proof that the real LSU will behave correctly since they
share identical logic. Note that these additional LSUs are used purely as lemmas to
help construct a proof about the original; the properties about them are not intended
to provide meaningful standalone results.

\subsection{Constraint on Capabilities coming from Memory}\label{src:memconstrain}

Unlike the original CHERI specification, the CHERIoT-Ibex specification
does not admit every sequence of bits as a representable and taggable
capability.  For example, it would not be possible to have a capability
with top bound greater than $2^{32}$. The details of this restriction are
important for our verification, because some of the microarchitectural
optimisations made by the designers of CHERIoT-Ibex are not sound without
this constraint.

In an ideal proof environment, an easy-to-verify assumption could be
written that restricts the capabilities read from memory to be only those
written earlier to memory.  But this would require a model of memory and
would be difficult to implement in practice.  We instead explicitly
constrain all loadable capabilities, with the constraint applied only to
capabilities that have their tag bit set. A faulty DMA device or additional
less well behaved cores may invalidate our asssumption, but that is beyond
the scope of our verification.

Finding a constraint sufficient to identify only well-formed capabilities
is not necessarily difficult.  But it is not immediately clear that any
restriction one might write down would be both minimal and true of all
capabilities that could be created without the use of memory.  We therefore
gain confidence in the correctness of the memory constraint by adding it to
our global data-type invariant. If the memory constraint always holds for
all tagged capabilities within the processor, then it always holds of
tagged capabilities written to memory---and hence it must be true for tagged
capabilities loaded from memory too.

From a methodological perspective, incorporating the memory constraint into
our global data-type invariant serves to validate the constraint as a
legitimate assumption about capabilities read from memory. Doing this also
has the added benefit of strengthening the DTI, which usually increases
both the speed of its own proofs and of dependent proofs. 

\section{Top-Level Observational Correctness}\label{sec:obs}

As outlined in Sec.~\ref{sec:methodology}, the top-level correctness
statement for CHERIoT-Ibex is that it produces the same stream of memory
interactions as the iterated Sail specification, when these are started in
the same initial state. In this section, we explain how this property is
established, given end-to-end instruction correctness
(Sec.~\ref{sec:end-to-end}) and memory correctness
(Sec.~\ref{sec:memory}). We explain the proof bottom-up.

\subsection{Continuity Properties}

In our approach to end-to-end instruction correctness, we step the pipeline
follower forward using internal processor signals, rather than some
independently-implemented control logic. This methodology is easy to
implement and powerful, and it finds many bugs. But it does not fully
verify the processor's pipeline control logic, which is often complex and
may contain subtle corner case bugs.

In our methodology, we therefore complement the end-to-end proofs with
certain \textit{continuity properties}. These assertions use the end-to-end
results as lemmas to prove that, in essence, each time the Sail
specification is `queried', the new inputs to the specification are
equivalent to (or at least as strict as, in the case of CHERI) the outputs
from the previous time the specification was queried.

To implement this approach, we introduce a small, self-contained piece of
SystemVerilog that, when the signal \verb|spec_en| is fired, will verify
that the new inputs to the compiled Sail specification---which are
abstracted from the microarchitectural state of the RTL---match the old
outputs. It also, in the same cycle, stores the new outputs to be checked on
the next iteration.

It is relatively easy to get conclusive proofs of the continuity
properties when the end-to-end correctness properties are provided as lemmas.
With some helper invariant work these proofs can be obtained in a matter of seconds
or minutes.

Successful verification of these properties extends the end-to-end properties to show
that not only does each instruction run correctly, but that the state we finish in is
the state the next instruction begins with.

\subsection{State Matching}

The next level of our verification establishes that a certain infinite
sequence of periodically sampled states of the processor matches the
sequence of states produced by iterating the Sail specification, when they
both start in the same initial state.  In particular, consider the infinite
sequence of microarchitectural states obtained by repeatedly sampling the
processor state when \verb|spec_en| is high.  Recall that this is the
signal that triggers an evaluation of the Sail specification.  State
matching says that the abstraction of this sequence of microarchitectural
states should match the sequence of architectural states produced by
iterating the specification.

Logically, it does not matter how \verb|spec_en| is defined, but in our
case it is high when either an instruction moves from the execute stage to
the writeback stage, or when an interrupt is handled.  It does, however,
matter that \verb|spec_en| is live. If \verb|spec_en| is not always
eventually fired then it may be possible for CHERIoT-Ibex to deviate from
the specification without ever being checked for correctness
again. Essentially, liveness for \verb|spec_en| ensures that states are
always eventually compared.

To prove this in our formal verification, we obtain a weak but finite upper
bound on the number of cycles between successive \verb|spec_en| states.
This is done under the assumption that a processor sleep following a WFI
will always be awoken from in bounded time, and that memory response times
are bounded, as explained earlier. We proved these weak bounds by
considering the `path' of microarchitectural states between one
\verb|spec_en| state and the next.  We then prove small bounds over the
edges of each state, which are then summed into longer paths until the full
cycle is reached.  The initial bounds can take a few minutes to prove, but
they compose more or less instantly into proofs for longer paths, including
\verb|spec_en| to \verb|spec_en|.

Having established liveness, we now provide an argument that the required
state matching property holds, given the formal SVA properties we have
established in the formal verification tool environment. We begin with some
notation.

\paragraph{Specification Notation}

First, it is helpful to introduce some mathematical notation to refer to
the results produced by iterating the SystemVerilog compilation of the Sail
specification. It is important to note that the meaning of this notation is
not defined mathematically or formally. It is instead an informal notation
for referring---in our argument that state matching holds---to what the
compiled specification code in fact does.

We write $S_A$ for the set of all \textit{architectural states}. These comprise
the values of all the registers defined in the SystemVerilog compilation of
the Sail specification, which going forward we will refer to simply as `the
Sail specification'. These are the global state of execution of the Sail
specification; in the compiled code, they are inputs and outputs of the
specification SystemVerilog module.

We now let $I$ denote the set of \textit{architectural inputs}. These
are all the non-state inputs to the Sail specification. They comprise the
instruction to be executed, the interrupts, and the responses that will
come from memory, in case the instruction input is a memory read ---
as discussed in Sec.~\ref{sec:memory}.

We can now introduce the following notation to refer to updates to the
architectural state that are produced by the Sail specification:
\[
\mathrm{Spec} : S_A \times I \to S_A
\]

\noindent The function $\mathrm{Spec}$ maps one architectural state to the
next. Essentially, it produces the specified mutation to architectural
state that the (compiled) Sail computes. This output is the next state that
the specification would be executed in if it were to be iterated.

\paragraph{Specification Strengthening}

It happens that CHERIoT-Ibex allow tags to be cleared in
some situations when the specification would not.  From a security
perspective, a limited deviation of this kind is reasonable and acceptable,
since the implemented behaviour is only stricter than the specification.
It does however mean that we cannot prove direct equivalence with the Sail
specification, but instead against a hypothetical,  stricter version of the
specification.

We first introduce a partial ordering $\sqsubseteq : S_A \times S_A$ on
architectural states. We say that for any $s, s' \in S_A$, 
$s \sqsubseteq s'$, exactly when  the tag bit of any capability in
a register of $s$ is set only if the tag bit of the capability in the same
register of $s'$ is also set, and all other state components are equal. We would then
say that $s$ is an architectural state that is `at least as strict' as $s'$.

We now introduce a function
\[
\mathrm{clear} : S_A \times I \times S_A \to S_A
\]

\noindent denoting the additional tag clearing that the CHERIoT-Ibex
implementation does. Given an architectural state and an architectural
input that are to be provided to the Sail specification, along with the next
architectural state that the specification would produce for these inputs,
$\mathrm{clear}$ produces the potentially more strict architectural state
that aligns with what the implementation would do:
\[
\forall s\in S_a, i \in I. \: \mathrm{clear}(s, i, \mathrm{Spec}(s, i)) \sqsubseteq
\mathrm{Spec}(s, i).
\]

\noindent That is, we will assume that $\mathrm{clear}$ will clear tag bits
of a state exactly when CHERIoT-Ibex would do the same. It is important to
note that $\mathrm{clear}$ is never explicitly defined in SystemVerilog in
our verification code. We introduce this notation solely to facilitate our
exposition of state matching. In practice, specification alignment is
implemented in the verification code in essence by implementing
$\sqsubseteq : S_A \times S_A$.

In what follows, we prove observational equivalence with respect to the idealised
specification $\mathrm{CSpec} : S_A \times I \to S_A$, defined by
\[
\mathrm{CSpec}(a, i) = \mathrm{clear}(a, i, \mathrm{Spec}(a, i)).
\]

\noindent Readers interested in the specifics of tag clearing in
CHERIoT-Ibex may refer to the \verb|set_address| function of the CHERIoT-Ibex
RTL~\cite{cheriot-ibex}.

\paragraph{Microarchitecture and State Mapping}

Mirroring the specification notation introduced above, we write $S_\mu$ for
the set of all \textit{microarchitectural states}. These are all possible
assignments of values to all the hardware registers in
CHERIoT-Ibex. Similarly, we write $I_\mu$ for the set of all
\textit{microarchitectural inputs}. These are the values present on the
input ports of CHERIoT-Ibex. As explained in Sec.~\ref{sec:methodology}, we
verify the instruction fetch part of the processor separately, so for our
purposes here the `input ports' include the instruction input signals delivered by
instruction fetch to the compressed instruction decoder.

Following decades of common practice in the processor verification
domain~\cite{aagaard}, we introduce an abstraction function $\mathrm{abs} :
S_\mu \to S_A$ that maps microarchitectural state to architectural state.

\paragraph{Temporal Alignment}

In our verification, the specification is purely combinational while the
implemented processor is pipelined, with instruction execution spread over
several clock cycles---including a dependence on the timing of memory
response. In our verification code, the implementation of the pipeline
follower is what achieves a temporal \textit{alignment}~\cite{aagaard} of the two levels. We 
now introduce further (informal) notation to refer to this.

We write $X^*$ and $X^\omega$ for the sets of finite and
infinite sequences of elements of $X$ respectively. If $x \in X^\omega$,
then $x_i$ refers to the $i$th element of $x$, where $x_0$ is the first
element.

Now define the set $T_\mu$ to represent the architectural timing
information for any given $i \in I$. An element of $T_\mu$ defines how long
memory operations will have to wait before receiving responses and for how
long the processor will have to stall before a new instruction is
received. $T_\mu$ is never represented explicitly in SystemVerilog; it
instead tracks the decisions made by the model to provide or not provide
some response at some time.

We introduce the notation $\mathrm{Realise} : I^\omega \times T_\mu^\omega \to I_\mu^\omega$
to denote the mapping from a sequence of architectural inputs, as
presented to the specification, to the corresponding sequence of
microarchitectural inputs presented to the CHERIoT-Ibex design over 
successive clock cycles, using the corresponding timing information.

It is important to note that there isn't a simple mapping from a single
architectural input to a contiguous finite sequence of microarchitectural
inputs that correspond it.  This is because a single invocation of the Sail
specification will, in general, mark the start of running CHERIoT-Ibex for
several clock cycles. During this period of time, due to the pipelined
nature of CHERIoT-Ibex, the microarchitectural inputs for several
instructions in flight may be sampled. For example, the memory responses,
provided symbolically to the specification as stated earlier, will
arrive at the memory input ports of the hardware in the expected clock
cycles for the memory read instruction that originated them,
even though these occur strictly after the clock cycle in which \verb|spec_en| next
holds, i.e. after the \verb|spec_en| state for the corresponding memory read
instruction, at which point another instruction may be running in the
previous pipeline stage with new instruction bits.

For the purposes of discussing state matching, the mathematical
notation `$\mathrm{Realise}$' we introduce here is an abstract representation of the
behaviour of certain SystemVerilog code in our machine-executed
verification, including elements of the pipeline follower. This code is
hand written and designed by reading the documentation for the CHERIoT-Ibex
memory ports. We note that $\mathrm{Realise}$ could easily be implemented
without access to any internal signals, but doing so achieves little more
than duplicating some small section of the implementation to measure the
same events. Hence for now we do not do this. We consider
$\mathrm{Realise}$ to be part of our top level correctness statement, as
it needs human interpretation to implement.

\paragraph{State Matching}

With this notation introduced, we may now prove the state matching
property.  We begin by letting $i \in I^\omega$ stand for an
arbitrary infinite sequence of architectural inputs, and $t \in T_\mu^\omega$
stand for some arbitrary infinite sequence of micro-architectural
stall timings. Note that in the verification code the entirety of $i$ and $t$ are, of course,
not constructed in advance. These sequences are generated lazily by the model
checker, as their elements are required.

We will say that the inputs provided to the specification for occurrence
$n$ of a specification query are just the architectural inputs $i_n$. Of
course, in real execution of the processor on concrete instructions, the
value of $i_n$ will be partially constructed in advance. For example, the
raw instruction bits will be determined several cycles in advance and
stored in the fetch FIFO before being presented to the (RVC) compressed instruction
decoder.

We now define $a \in S_A^\omega$ to be the infinite sequence of
architectural states that arise from $i$. We suppose that $a_0$ is the initial
architectural state of the specification. The infinite sequence of
subsequent architectural states is then defined by 
\[
a_{n{+}1} =
\mathrm{CSpec}(a_n, i_n)\quad \textrm{for all $n \ge 0$.}
\]

We now let $i' = \mathrm{Realise}(i, t)$ be the infinite sequence of
microarchitectural inputs sent to CHERIoT-Ibex, one for each clock cycle,
that corresponds to the sequence $i$ under timings $t$. We define $s \in S_{\mu}^\omega$
to be the infinite sequence of microarchitectural states that arise from
$i'$, with $s_0$ being the reset state of the microarchitecture.  We
argued above that \verb|spec_en| will be true infinitely often, under any
sequence of inputs to CHERIoT-Ibex. So we can define $\mathrm{en}(s)\in
S_{\mu}^\omega$ to be the infinite subsequence of $s$ obtained by sampling
the sequence $s$ whenever \verb|spec_en| is true, where
$\mathrm{en}(s)_0$ is the microarchitectural state when \verb|spec_en| is
true for the first time. We use the notation $\mathrm{en}(s)_{-1}$ to refer to the
microarchitectural state immediately preceding the reset state.

Finally, the state matching property says that for all $n \geq 0$,
\begin{equation}
\mathrm{abs}(\mathrm{en}(s)_n) = a_{n}.
\label{state-matching-thm}
\end{equation}

\noindent That is, every time the specification module in our verification
code is invoked on an abstraction of the current microarchitectural
state---i.e.\ exactly when \verb|spec_en| is high---the abstracted architectural
state produced in our verification code matches the iterated architectural state
of the specification. The proof proceeds by induction on $n$, as follows.

For the base case, when $n$ is $0$, then \verb|spec_en| is high for the
first time. We take $\mathrm{abs}(\mathrm{en}(s)_0) = a_0$ as an as of yet
unproven assumption.

Now suppose that our induction hypothesis (\ref{state-matching-thm}) holds
for an arbitrary $n \geq 0$ and consider microarchitectural state
$\mathrm{en}(s)_{n{+}1}$.  At the previous clock cycle when
\verb|spec_en| was high (in state $\mathrm{en}(s)_{n}$), we stored the output
of the implemented specification module:
\[
\mathrm{CSpec}(\mathrm{abs}(\mathrm{en})_{n}, i_n) = \mathrm{CSpec}(a_{n}, i_n) = a_{n{+}1}
\]
By the inductive hypothesis. Our SVA continuity properties directly assert that
$\mathrm{abs}(\mathrm{en}(s)_{n{+}1})$ is equal to this stored state. So we
have that  
\[
\mathrm{abs}(\mathrm{en}(s)_{n{+}1}) = a_{n{+}1}
\]
\noindent as required.

\subsection{Observational Correctness}

From state matching (\ref{state-matching-thm}) we can conclude that under a
`reasonable' definition of $\mathrm{abs}$ the sequence of internal states
of CHERIoT-Ibex will match those of the specification infinitely
often---specifically, whenever \verb|spec_en| is high.  But this is not
quite abstract enough for our final statement of correctness. This is
because the significance for correctness of this property depends on the
definition of $\mathrm{abs}$, which is derived from internal signals of
CHERIoT-Ibex. Observational correctness improves on this by instead
formulating correctness by comparing only outputs. We sketch the argument
below.

\paragraph{Notation}

We will write $O$ for the set of all \textit{architectural outputs}. These
are the outputs of the Sail specification that represent the values
involved in memory events. In the case of a memory write, the values are
the address, the data sent to memory, and certain byte enable flags. In
the case of a memory read, we have only the address and byte enable
flags. Finally, there is a value to represent `no memory event'. We 
can then introduce the notation
\[
\mathrm{SpecOut} : S_A \times I \to O
\]

\noindent to stand for a mapping from the current architectural state and
inputs to the outputs that the specification produces.

For the implementation level, we introduce $O_\mu$ for the set
of all \textit{microarchitectural outputs}. These are simply the values on
the memory output ports of CHERIoT-Ibex on each clock cycle. These values
comprise the memory request, write enable and byte enable signals, and any
data that is to be written to memory. The correspondence between values in
$O_\mu$ and values in $O$ is straightforwardly implemented in our
SystemVerilog code.

\paragraph{Observational Correctness}
Suppose $i\in I^\omega$ is an arbitrary infinite
sequence of architectural inputs and $a\in S^\omega_A$ is the infinite
sequence of architectural states that arise from $i$. We can introduce
$o \in O^\omega$ defined by
\[
o_n = \mathrm{SpecOut}(a_n, i_n) \quad \textrm{for all $n \ge 0$}
\] 
\noindent to stand for the corresponding infinite sequence of memory events
that would be obtained by hypothetically iterating the specification alone.

We further define $\mathrm{Check} : O \times T_\mu \to O_\mu^*$ to stand for a function
that maps each architectural output in the sequence $o$ to the
corresponding \textit{finite} sequence of microarchitectural outputs that
would have to be generated by the CHERIoT-Ibex core to correctly enact this
architectural output. This depends on the timing information for that architectural
cycle, since requests sent to memory will only be granted after some micro-architecturally
defined time.

As with $\mathrm{Realise}$, the mathematical notation `$\mathrm{Check}$' is
an abstract representation of some SystemVerilog code in our machine-executed formal
verification, and we regard this function as part of our top level statement,
needing human involvement to implement.

The top level memory properties verify directly in SVA that for all $n \ge 0$ and
any $t\in T_\mu^\omega$, if $o'_n \in O_\mu^*$ is the
sequence of memory outputs produced by CHERIoT-Ibex as it
progresses from state $\mathrm{en}(s)_{n{-}1}$ (exclusive) to $\mathrm{en}(s)_{n}$
(inclusive) under timings $t_n$, then
\begin{equation}
\mathrm{Check}(\mathrm{SpecOut}(\mathrm{abs}(\mathrm{en}(s)_{n}), i_n), t_n) = o'_n.
\label{top-level-mem}
\end{equation}
\noindent Given (\ref{top-level-mem}), we may now apply state matching (\ref{state-matching-thm}) to obtain
\[
\mathrm{Check}(o_n, t_n) = o'_n.
\]
\noindent Since there are infinite \verb|spec_en| states, we can then conclude
that CHERIoT-Ibex and the tightened specification will produce precisely
corresponding outputs forever, up to micro-architecturally defined timings and
under suitable implementations of $\mathrm{Realise}$ and $\mathrm{Check}$. This is the highest level
statement one could hope to make in this context, and establishes
observational equivalence.

\paragraph{Top Level Memory Properties}

We now explain how the top level memory properties prove ($\ref{top-level-mem}$).
First, we note that our formal verification properties include a check
that, for the case of memory instructions, the memory outputs from the compiled
specification module in SystemVerilog remain stable during the period from
the clock cycle at which CHERIoT-Ibex begins its first request (if any),
up to and including the clock cycle in which the next \verb|spec_en| state occurs. It
is therefore legitimate to take this stable output result from the compiled
specification as the unique reference for correctness of the machine's
outputs during this period.

Now, for each clock cycle of the relevant sequence of
microarchitectural states, our memory correctness properties check that if
a memory request is made, it is correct with respect to the sequence
stipulated by $\mathrm{Check}(o_n, t_n)$. Correctness here means that in each cycle
in which a memory request is in fact made by the processor, the addresses,
byte flag, and (in the case of memory writes) data match what the reference
sequence requires. Our memory properties further assert that the same
number of memory requests are made by the CHERIoT-Ibex core as $\mathrm{Check}(o_n, t_n)$
requires. Combined, this means that if CHERIoT-Ibex makes a request it does so correctly,
and that if it is expected to make a request is makes one.

\section{Design Bugs Revealed}

\begin{figure*}[bt]\small
\centering
\begin{center}
\rowcolors{1}{white}{gray!25}
\begin{tabular}{ l | l }
  Illegal CLC load & CLC tag bit leak \\
  CSeal otypes & CJALR alignment checks \\
  CSEQX memory vs.~decoded & MTVEC/MEPC legalisation (in several instances) \\
  CSC alignment checks & CSC decoding \\
  Store local violation & Memory capability layout \\
  PCC.address vs.~PC & CJAL vs.~CJALR \\
  Memory bounds check overflow & CLC tag/perms clearing \\
  MSHWM/MSHWMB updates & \verb|tvec_addr| alignment \\
  Sealed PCC & CSetBounds lower bound check \\
  Illegal instruction MTVAL values & Memory and branch exception priorities \\
  CSpecialRW exception priorities and SLC issues & EBreak MTVAL values \\
  MRet MStatus.MPRV & 16 vs.~32 register spec issues \\
  IF granules and overflow & MEPCC \verb|set_address| \\
  User mode WFI & CSR instruction problems \\
  Undocumented CJALR & PMP pipeline flushing on CSR clear \\
  TRVK RF write collision & Stack\_EPC for CHERI NMIs \\
\end{tabular}
\end{center}
\caption{List of bugs found in CHERIoT-Ibex~\cite{CHERIoT-Ibex-issues}. The PMP (and arguable NMI) bugs are relevant only to vanilla Ibex.}\label{bugs.fig}
\end{figure*}

Our work revealed around 30 bugs in the CHERIoT-Ibex design---virtually all
to do with the capability extensions added to the original Ibex core. These
were reported to the design team, confirmed, and fixed or remained under
discussion at the time of writing~\cite{CHERIoT-Ibex-issues}. The bugs found range from minor
inconsistencies to exploitable, monotonicity-breaking vulnerabilities. We
focus on the latter here, as these are of primary interest in a processor
designed as a foundation for system security.

This bug count is a conservative lower bound. We have, for example, grouped
a number of exception priority anomalies into one `bug', and don't count
new, related bugs introduced by trying to fix an already found bug. We also
note that the design was at a relatively early state of maturity when we
began formal verification. We have no doubt that some of the bugs revealed
by formal would have surfaced under intensive design simulation or Burch-Dill
flushing~\cite{10.1007/3-540-58179-0_44} type formal verification. However many
of the more interesting corner case bugs are unlikely to have
been uncovered without multiple instructions interacting with one another.

It is notable that the new Sail to SystemVerilog flow enabled us to get verification underway
fast and to start finding bugs very quickly. Once a pipeline follower is set up
bug hunting may begin immediately under bounded verification. While we have gone to great
lengths to obtain fast converging proofs for our verification, successful bug hunting,
especially in the early stages of developement, does not strictly require it.

The following account of some of the bugs we found is somewhat technical
and makes significant reference to the specifics of CHERI semantics. For a
full understanding of the details, it may be useful to read in conjunction
with the main reference for compressed capabilities~\cite{CHERIConcentrate},
the CHERIoT-Ibex semantics~\cite{cheriot-sail}, and the Microsoft
CHERIoT technical report~\cite{amar2023cheriot}.

The first vulnerability was discovered by the DTI. It allows for the
moving of the address of a capability without doing an associated representability
check. Such checks are made whenever the address of a capability is changed.
Their purpose is to ensure that the new bounds are identical to the old ones (since those
bounds are a function of the address). If the new bounds are different the tag of the offending
capability would be cleared. A missing representability check
will be caught by the DTI since the \verb|top_cor| and \verb|base_cor| values
should also be recomputed when the address is changed.

The CLoadCapImm instruction loads a capability from memory by dereferencing
another `authorising' capability.  The bug arose when a specific bit of a
CLoadCapImm instruction was set which should render the instruction illegal.
The instruction would decode regardless, and yet it would still be marked as illegal.
This caused an exception to be raised, but the load command would nonetheless continue to
be dispatched to memory. This is a bug, and the fix was to prevent that load
command being sent.

Without the fix however, if the response from memory took long
enough to arrive, the load-store unit could return the result to the writeback
stage at the same time that another instruction was retiring. By a simple
and reasonable design choice of the writeback stage, the addresses of the
illegally loaded capability and of the capability coming from the other
instruction would be ORed-together before being sent to the register file for
storage. This means that the address of the capability changed without any
representability check being made, meaning the bounds could move.

An overflow was also found during instruction checking, affecting the bounds
checking for all memory loads and stores. In particular, a memory operation
would fail if $a <_\mathrm{unsigned} t - w$, where $a$ is the requested
address, $t$ is the top field, and $w$ is the requested width.  Then if $t
= 0$ and $w = 4$, the check becomes $a <_\mathrm{unsigned} 2^{32} - 4$.  This
meant that accesses could be made outside of the range of the given
capability, clearly breaking monotonicity. The fix took several iterations to
finalise and was developed in close discussion between the developer and our
verification team~\cite{issue12}. This, we believe, is the monotonicity bug also
documented in~\cite{vericheri2024}.

A further, somewhat subtle bug was found in memory permissions clearing. When a
capability is loaded from memory, it will lose some of its permissions
based on the permissions of the authorising capability. For instance, if
the authorising capability was not marked as mutable, then the loaded
capability would be stripped of its mutability too. Since capability
permissions are compressed and not all combinations are represented, it is
possible that removing a permission will result in a set of permissions
that is not representable. The specification therefore requires that the
largest representable subset of permissions should be selected when
permissions are being refined. Due to a small error in the RTL, and a quirk
in the encoding scheme for permissions, the clearing of a certain bit
resulted in greater permissions than were originally had, thus again
breaking monotonicity.

The CSetBounds instruction is intended to narrow the bounds of a
capability.  It updates the lower bound of a capability to be the address
of that same capability and the length of the capability to be a value
given in a separate register. If the new length
meant that the new top would exceed the previous top, then the tag
of the new capability was correctly cleared.  The tag was however not
cleared if the address of the capability was less than the existing base.
This demonstrates the
importance of detailed verification of the invariants of the processor,
since it was not necessarily clear that such a capability could have been
created by the CSetAddress family of instructions without losing its tag
bit.

The final monotonicity-breaking bug was also found by the DTI. In RISC-V,
when an exception or interrupt occurs, the address of the running/next to
run instruction is saved to the MEPC CSR. For CHERI RISC-V, this is
extended to the whole PCC that is saved in MEPCC,
but where the PC is still that of the instruction on which the exception or
interrupt occurred.   This requires changing the address of the PCC,
so this would ordinarily need a representability check, as in the first
bug. In this case, however, the designers believed that none was needed, since
instruction fetch should have already dealt with the possibility the PC
was outside of the bounds of the PCC.

This invariant was found to be false. If a branch was made to a location
outside of the bounds of the PCC, the PC would be updated and a request to
instruction memory would be sent.  No exception would be raised until
memory responded, so suppose it took several cycles to do so.  If a
hardware interrupt was raised in the time it took for that response to
come, the processor would go to update MEPCC. In particular, it would use
the PCC with the new out of bounds PC, without checking representability
and thus breaking monotonicity again.

This was caught by the memory constraining part of the DTI, since this resulted in a capability that
had a top value above $2^{32}$, which should be impossible.

Two bugs were also found affecting vanilla Ibex. One was discovered
accidentally early on and was a decoding issue found in the raising of
illegal instruction exceptions in E~mode. The bug is more or less harmless.
The more interesting bug affected instruction fetch PMP checks. In
particular, if a CSR clear instruction was used to mutate some PMP registers,
then the pipeline would not be flushed, and the PMP check for the
instruction bits of the next instruction would be done with the old PMP
registers, not the new.
It is unclear how likely changing PMP registers with CSR clear instructions is.
However, to preserve the PMP protection model compilers must flush the pipeline or otherwise ensure that the PMP change takes effect before doing a memory access using the affected memory regions.
This is clearly not a significant vulnerability,
but it does demonstrate that our methodology is capable of discovering bugs
in even production-ready code, with extensive functional verification~\cite{ibex-verification}.

\section{Verification Coverage}\label{sec:coverage}

Our formal verification covers the default configuration of CHERIoT-Ibex,
with CHERIoTEn, RV32E, RV32BNone, in pure capability mode, with temporal safety.
Memory operations are assumed to respond to requests within a fixed number of
cycles and there is no PMP for CHERIoT-Ibex.  Interrupts are assumed to remain active until
the interrupt is cleared by a memory operation. It is further assumed, for
liveness, that an interrupt arrives within a bounded amount of time
following execution of WFI.

Configuring \verb|RV32B = RV32BNone| disables the bit manipulation extension
to RISC-V, meaning the instructions it brings are not present in the version
of CHERIoT-Ibex we verified. Other formal verification efforts by another group have
revealed a bug affecting some instructions from that extension~\cite{CHERIoT-Ibex-issues}.

The implementation has some additional features the Sail does not specify,
meaning those behaviours cannot be verified. In particular, we forbid
attempting to enter debug mode and memory bus errors cannot be raised. Some CSRs, such as
debug CSRs, MSHWM/MSHWMB and performance counters are not checked, this is largely
also due to missing or incomplete specifications.
We expect that these features can be tested using a traditional functional verification approach.

The vanilla Ibex verification additionally includes checking correctness
of the PMP implementation (and in particular Smepmp), which increases
complexity, though in exchange all of the CHERI complexity is lost.
Our results are otherwise the same as for CHERIoT-Ibex.

Everything the specification or the implementation can do is verified in all cases.
This includes all instructions, interrupts, fetch errors, exceptions and anything
else, with the exception of the special cases mentioned above and of M-Type instructions.
These are arithmetic data path instructions for which we did not attempt to find
conclusive proofs of correctness. We expect that these will be
provable with the aid of state-of-the-art data path verification
technology, such as symbolic trajectory evaluation~\cite{Seger:2005:IEE}.

When an unknown CSR is accessed, the relevant end-to-end properties
are not checked and the \verb|spec_en| step from Sec.~\ref{sec:obs}
is left unchecked too, though the specification outputs are still stored.
Everything else mentioned above that we do not check is simply assumed not to
occur.

Finally, the correctness of instruction fetch was also proved. The Ibex
instruction fetch module is essentially a FIFO that stores responses from
instruction memory. Since there is no real equivalent in the Sail, this
proof is done outside of main verification for observational correctness. We
prove that if an instruction enters that FIFO with a given address, then if
it later leaves the FIFO, it must do so with the same address.

\section{Related Work}\label{sec:related}

The literature on processor verification is, of course, voluminous by now
and documents many valuable experiments and contributions. We will not
include an extended survey in this report, but focus here on only the most
closely related works that address the verification of CHERI processors
specifically or the central end-to-end component of our methodology.

The first substantial verification effort for a CHERI processor was
reported by Gao and Melham in~\cite{Gao:2021:EFV}. This targeted
CHERI-Flute~\cite{CTSRDCHERIFluteRISCV}, a modified version of Bluespec's
Flute RISC-V processor~\cite{Flute,FluteAnnouncement} that implements CHERI
RISC-V instructions. Flute is an open-source 64-bit RISC-V processor with a
five-stage, in-order pipeline---so along some dimensions CHERI-Flute is
larger in scale than CHERIoT-Ibex. Capabilities in CHERI-Flute are
decompressed and compressed when they are moved between memory and the
core. Internally, in registers, they are held in decompressed form.

The main result of this work was the end-to-end verification of all the
implemented CHERI instructions in the processor, together with certain
processor liveness results.  Several bugs were revealed by this
verification effort.

The specifications were manually translated from Sail to SystemVerilog, and
the effort needed for this is reported as having limited the coverage of
the specifications. Only instruction executions that do not lead to a
failure were verified, because the manual effort needed to specify the
correct outcomes following the `more than a dozen ways' in which an
instruction can fail was prohibitive given the (substantial) time available.
The lack of a flow from Sail to the verification tool is cited
in~\cite{Gao:2021:EFV} as the practical limiting factor, and this
experience inspired us to create the Sail to SystemVerilog flow reported
here.

Another difference is the proof engineering methodology used
for the end-to-end correctness properties. To achieve conclusive proofs,
Gao and Melham employed a strategy they call `decomposing the pipeline', in
which a chain of lemmas are formulated that track the required state
updates back through the pipeline and allow $k$-induction to converge. This
required thorough analysis of pipeline forwarding and other
microarchitectural details. By contrast, the approach developed here relies
on more ad-hoc, but methodologically formulaic, invariant
development.

\medskip

The other published work to date that targets CHERIoT-Ibex specifically is
VeriCHERI~\cite{vericheri2024}.  This approach is complementary to our
functional verification research; VeriCHERI is a `specification free'
approach that directly encodes desirable properties about the security of
memory accesses and seeks to verify that the RTL conforms to these
properties. In essence, the properties state that no memory transaction is
issued whose address lies within a symbolically-enumerated `protected'
region of memory, assuming that no capability that legitimately addresses
this region is present in any register of the processor. A `weak
monotonicity' property is also established, which states that this
assumption is maintained across a single step of the processor---and hence
by induction that it is an invariant.

Research demonstrating that security-related properties are guaranteed by a
CHERI ISA includes proofs that such properties follow from the Sail
specification~\cite{cheriot-sail-properties}.  Observational correctness
with respect to a Sail specification, as we demonstrate here, implies that
the processor RTL also satisfies these properties.
However, this approach does not detect side channels that could be
exploited by a timing attack. VeriCHERI, however, can reveal
microarchitectural side channels. This is demonstrated
in~\cite{vericheri2024}, which documents a Meltdown-style vulnerability in
CHERIoT-Ibex by which a timing attack could, in principle, gain a small
amount of information about an arbitrary word in memory. This vulnerability
cannot, of course, be found with a functional verification of the kind we
report here.

\medskip

A landmark paper on end-to-end instruction verification is the work of Reid
et al.~on ISA-Formal at Arm~\cite{Reid-2016-EEV}. This was the inspiration
for our starting point in the CHERI-Flute verification
effort~\cite{Gao:2021:EFV}.  There are many similarities to the
end-to-end instruction verification component of the methodology we
illustrate here with CHERIoT-Ibex.  This, however, is only one part of our
overall proof that establishes top-level observational correctness.

In matters of detail, our approach to specification installation using a
pipeline follower departs from that explained
in~\cite{Reid-2016-EEV}, especially with regards to memory. Where we
can make use of the predictable and linear ordering of memory requests
of CHERIoT-Ibex, the larger ARM chips verified under ISA-Formal make weaker
guarantees, with some instructions (e.g. LDM and STM) making up to 16 memory requests in
no particular order. For this reason in ISA-Formal memory operations were
collated and verified at the end of their pipeline stage, instead of as they
occurred as in our work.

Additionally, in the ISA-Formal work, at least for simple cases, the
architectural pre-state is sampled (through an abstraction function, as here)
at the time of retirement from the writeback stage. This is compared (in the same
cycle) to the specification computed on the architectural post-state derived
from the outputs of the memory stage. This differs from our model of carrying
the specification post-state in the pipeline follower up to the point of retirement.
ISA-Formal uses the pipeline follower primarily to store the running instruction
opcode. The explanation~\cite{Reid-2016-EEV} is
by illustration with a simple register-to-register addition function, and there
may well be variations to this structure for other instruction classes.

Analogous to the Sail flow reported here, this is based on a compilation
from specifications expressed in Arm's Architecture Specification Language
to SystemVerilog. Here, too, one of the challenges was bitstring
polymorphism, necessitating a monomorphisation pass of the compiler.

Of course a significant difference in practice between ISA-Formal and our
work is that the former targets Arm processors of much greater complexity
than a RISC-V core---at least the one that CHERIoT-Ibex is based on. The
Arm verifications, therefore, obtained only bounded proofs, and
did not cover exceptions and `the memory management unit'.  Nonetheless,
bounded proofs and partial verification coverage can be highly effective at
bug-finding and can deliver significant value in practice.

\section{Concluding Remarks}

In this paper, we present the first comprehensive formal verification of a
CHERI RISC-V processor with internal
compressed capabilities, against an authoritative specification in the Sail
ISA specification language. This was made possible by the development of a
new Sail to SystemVerilog flow, and by extensive methodological and
verification-structuring work. We hope that this account of our work will be
a useful guide to verification engineers embarking on a similar project.

Our methodology makes considerable use of problem decomposition, employing
case splitting, extensive assume-guarantee reasoning, and $k$-induction,
executed entirely within a model-checking environment. These kinds of proof
strategies are used extensively by advanced practitioners of formal
verification by model checking, but they will also be very familiar to an
experienced user of deductive theorem provers, such as HOL~\cite{Gordon:1993:ITH} or
Lean~\cite{lean}.

Special attention was paid to making the proofs not only converge but run
fast. Both goals were typically pursued by analysing counterexamples from
failed proofs by $k$-induction to understand and characterise targeted
subsets of unreachable states. We could then formulate and prove internal
invariants that would make the proofs converge or run fast. This was done
manually in this project, and it is straightforward enough for a
verification engineer who understands microarchitecture and can formulate
logical invariants. Our current work is looking at automation support to
increase efficiency.

The compiler is in an early and experimental stage, and while it does work
well from a verification standpoint, it has several drawbacks from a larger
usability perspective. There are two issues in particular.  The first is
that the elaborated size of the specification is reasonably large. We intend to
investigate the use of DAG inlining~\cite{dag-inlining} to help deal with this.
Improvements to the way the specification handles some control flow would likely
also be of use.

The second and probably larger problem is that counterexamples can be
difficult to debug. The tools we used are prevented from providing
meaningful insight into the internal workings of the compiled specification
due to the general unreadability of the machine-generated code and the
heavy use of procedural code and SystemVerilog functions.  Both issues can
be resolved with improved tooling, either with dedicated debugging tools or
back-annotation and improvements to the compiler.

A further enhancement could be to generate abstraction points within the
specification itself. This would likely help significantly with proof
times, since instructions not actively being referred to could be
abstracted out entirely.

\section{Acknowledgements}

This research was funded in part by the UKRI programme on Digital Security
by Design (Ref.~EP/V000225/1, SCorCH~\cite{SCorCH}) and the Innovate UK
project Digital Security by Design (DSbD) Technology Platform Prototype,
105694.  This project has also received funding from the ERC under the
European Union's Horizon 2020 research and innovation programme (grant
agreement No.\ 789108, AdG ELVER, Sewell). Significant developments to the
verification for Ibex, ported also to CHERIoT-Ibex, were undertaken by the
first author while a summer intern at lowRISC. We are grateful to Marno van
der Maas of lowRISC for technical discussions and helpful editorial
contributions to drafts of this paper.

We are grateful to Alastair Reid (Intel) and Laurent Arditi (Codasip) for
insightful discussion of the best way to translate Sail into Verilog for
formal verification. Alastair also provided valuable feedback on the
final draft of this paper. Vincent Reynolds (Cadence) kindly provided
access to some draft technology for analysing a very deep counterexample
trace that had puzzled us for some time. We thank Peter Sewell (Cambridge)
for discussions and his support.

We are immensely grateful to the entirety of the CHERIoT team (Microsoft)
for their assistance and advice. We are particularly grateful to Kunyan Liu
for working through the issues we found (especially the mundane ones), to
Robert Norton-Wright for helping us identify invariants we had difficulties
finding ourselves, to David Chisnall for his deep insight into the nature
and history of the Sail specification, and to Wes Filardo for his input
from perspective of the CHERIoT-RTOS.

Grateful thanks are also due to Ohad Kammar, for his thoughtful review of
our top-level formulation of correctness. 

\bibliographystyle{IEEEtran}
\bibliography{Ploix-2025-CFV}

\typeout{get arXiv to do 4 passes: Label(s) may have changed. Rerun}

\end{document}